\documentclass[aps,pra,onecolumn,showpacs,amsmath,superscriptaddress]{revtex4}
\usepackage{xspace,epsfig,amsfonts,amssymb}
\usepackage{graphics,amstext,color}

\newcommand{\beq}{\begin{equation}} 
\newcommand{\eeq}{\end{equation}} 

\newcommand{\ba}{\begin{array}} 
\newcommand{\ea}{\end{array}} 

\def\beqa{\begin{eqnarray}}
\def\eeqa{\end{eqnarray}}

\newcommand{\ket}[1]{\vert \mkern1mu {#1} \mkern1mu \rangle} 
\newcommand{\bra}[1]{\langle \mkern1mu {#1} \mkern1mu \vert} 
\newcommand{\bracket}[2]{\langle \mkern1mu {#1} \mkern1mu \vert \mkern1mu {#2} \mkern1mu \rangle} 

\renewcommand{\mod}{\text{ mod }}

\newcommand{\arrow}[1]{\stackrel{#1}{\longrightarrow}}

\newcommand{\longarrow}[1]{\stackrel{#1}{\text{\hbox to 4.5em{\rightarrowfill}}}}

\font\bm=cmmib10

\begin{document}

\baselineskip 13pt

\title{Deterministic Quantum Distribution of a {\bm d}\,-ary key}





\author{Anita Eusebi}
\email{anita.eusebi(at)unicam.it}

\affiliation{Dipartimento di Matematica ed Informatica, Universit\`{a} di Camerino, 
I-62032 Camerino, Italy}

\author{Stefano Mancini}
\email{stefano.mancini(at)unicam.it}

\affiliation{Dipartimento di Fisica, Universit\`{a} di Camerino, 
I-62032 Camerino, Italy}

\begin{abstract} 
We present an extension to a $d$-ary alphabet of a recently proposed deterministic 
quantum key distribution protocol. 
It relies on the use of mutually unbiased bases in prime power dimension $d$, 
for which we provide an explicit expression.
Then, by considering a powerful individual attack, we show that the security of 
the protocol is maximal for $d=3$.
\end{abstract}


\maketitle


\section{Introduction}

Quantum Key Distribution (QKD) is recognized to complement the One Time Pad 
to a secure system for reliable transfer of confidential information 
\cite{PW98}. 
A paradigm for QKD (not exploiting entanglement) is the pioneering BB84 
protocol \cite{BB84}. 
It allows two remote parties (Alice and Bob) to share a secret key by a 
\emph{unidirectional} use of a quantum channel 
(supplemented by a public authenticated classical channel).

Protocols like BB84 have a \emph{probabilistic} character, in the sense that, 
on each use of the quantum channel, the sender (Alice) is not sure that the 
encoded symbol will be correctly decoded by the receiver (Bob).
Tipically, this only happens with probability $1/2$.

Recently a new generation of protocols has been introduced making the QKD 
process \emph{deterministic} \cite{BEKW02, BF02, CL04, LM05}.
In this case Alice is sure about the fact that Bob will exactly decode the 
symbol she has encoded.
This paradigm shift has been realized by a \emph{bidirectional} use of the 
quantum channel.
These new generation protocols are more versatile than the old 
generation ones and are supposed to outperform them. 

As much as like extensions of BB84 to larger alphabets have been developed 
\cite{BT99,CBKG01}, there is a persistent aim to also extend the protocol 
of \cite{LM05} to larger alphabets, that is to higher dimensions.
A construction has been recently devised for a tri-dimensional alphabet 
\cite{SLW06, SW07}, and then another for a continuous infinite-dimensional 
alphabet \cite{PMBL06}.

Here we present a protocol that realizes an extension of the deterministic 
protocol of \cite{LM05} to a $d$-ary alphabet. 
Since our construction is based on Mutually Unbiased Bases (MUB) 
\cite{I81, WF89, BBRV01, KR03}, it holds only for prime power dimensions $d$.
We will provide an explicit expression for MUB encompassing powers of both 
even and odd primes, by correcting the one given in \cite{D05}.
 
We then consider a powerful individual attack on the forward and 
backward path of the quantum channel and we show that the security 
for $d=3,4,5$ is higher than that at $d=2$ and is maximal for $d=3$.


\section{Qudits and Mutually Unbiased Bases}

Let us consider a qudit, i.e., a $d$-dimensional quantum system, and indicate with 
$\mathcal{H}_{d}$ the associated Hilbert space.
A set of orthonormal bases in $\mathcal{H}_{d}$ is called a set of 
\emph{Mutually Unbiased Bases} (MUB) if the absolute value of the inner product of 
any two vectors from different bases is $1/\sqrt{d}$ \cite{I81, WF89, BBRV01, KR03}. 

It is known that in $\mathcal{H}_{d}$, when $d$ is prime power, there exists 
a maximal set of $d+1$ MUB \cite{I81, WF89, BBRV01, KR03}.
Here, we focus on this case. 

From now on we assume that $d=p^m$, with $p$ a prime number and $m$ positive 
integer, and we denote the $d+1$ MUB of $
\mathcal{H}_{d}$ 
by $\ket{v_t^{k}}$, with $k = 0, 1, \ldots, d$ and $t = 0, 1, \ldots, d-1$ 
labelling the basis and the vector in it respectively.

Thus, for every $k, k' = 0, 1, \ldots, d$ and every $t, t' = 0, 1, \ldots, d-1$, 
the following equality holds:
	\beq \label{mub}
\left|
\strut\smash{\bracket{v_t^{k}}{v_{t'}^{k'}}}
\right|
=\frac{1}{\sqrt{d}}\left(1-\delta_{k,k'}\right)
+\delta_{t,t'}\delta_{k,k'},
	\eeq
where $\delta$ stands for the Kronecker delta.
 
We deal with the Galois field $G=\mathbb{F}(p^m)$ of $d$ elements.
We denote by $\oplus$ and $\odot$ respectively the addition and 
the multiplication in the field $G$ 
(by $\ominus$ and $\oslash$ the subtraction and the division in $G$). 
Usually, an element of $G$ is represented by a $m$-tuple 
$(g_0, g_1, \ldots, g_{m-1})$ of integers modulo $p$.
According to this representation, $\oplus$ corresponds to the componentwise 
addition modulo $p$.

Following \cite{D05}, we identify $G$ with $\{0, 1, \ldots, d-1\}$, paying 
attention to distinguish the operations in the field from the usual ones.
Namely, we identify $(g_0, g_1, \ldots, g_{m-1})$ with the integer 
$g=\sum_{n=0}^{m-1}g_{n}p^{n}$.
This allows us to consider the vector label $t$ in $\ket{v_t^{k}}$ as an 
element of $G$.

Let us denote the $p$-th root of unity by
	\beq
\omega = e^{i2\pi /p}.
	\eeq
It is proved in \cite{D05} that 
	\beq \label{OperExp_G}
\omega^{j} \cdot \omega^{l} = \omega^{j \oplus l} 
\quad \textnormal{with } j, l \in G 
	\eeq
and
	\beq \label{la4}
\sum_{j=0}^{d-1} \omega^{j \odot l} = d\,\delta_{l,0}
\quad \textnormal{with } l \in G.
	\eeq

We choose $\{\ket{v_t^0}\}_{t=0, \ldots, d-1}$ as the computational basis and 
use the explicit formula given in \cite{D05} to express the vectors of any 
other basis in the following compact way: 
	\beq \label{ket_d}
\ket{v^k_t} = {1 \over \sqrt d} 
\sum_{q=0}^{d-1}
\omega^{\ominus q \odot t}
(\omega^{(k-1) \odot q \odot q})^{\frac{1}{2}}
\ket{v^0_q},
	\eeq
where  $k = 1, \ldots, d$ and $t = 0, 1, \ldots, d-1$. In particular for $k=1$:
	\beq
\ket{v^1_t} = {1 \over \sqrt d} 
\sum_{q=0}^{d-1}
\omega^{\ominus q \odot t}
\ket{v^0_q}.
	\eeq

As it is pointed out in \cite{D05}, for $p$ odd the square root coincides with 
the division of the exponent by 2 in $G$ and it is uniquely determined.
On the contrary, for $p=2$ it is necessary to unambiguosly determine the square 
root's sign. This is given by (see Appendix)
	\beq \label{srs}
(\omega^{(j-1) \odot q \odot q})^{\frac{1}{2}} = 
\prod_{\textstyle{n=0 \atop q_n \neq 0}}^{m-1} 
\! i^{(j-1)\odot 2^n \odot 2^n} 
\omega^{(j-1) \odot 2^n \odot (q \mod 2^n)}. 
	\eeq
With this in mind, the expression (\ref{ket_d}) satisfies the condition (\ref{mub})
of MUB, for $d$ any prime power, both even and odd (see Appendix for the proof). 
Notice that this does not happen in \cite{D05} in the even case. Hence, in the 
following we will make use of (\ref{ket_d}) without distinguishing the two cases.


\section{The protocol}

Moving from the protocol of \cite{LM05}, we consider Bob sending to Alice a qudit 
state randomly chosen from the set 
$\{\ket{v_{t}^{k}}\}^{k = 1, \ldots, d}_{t = 0, \ldots, d-1}$ of MUB. 
Then, whatever is the state, Alice has to encode a symbol belonging to a $d$-ary 
alphabet $A = \{0, \ldots, d-1\}$ in such a way that Bob will be able to 
unambiguously decode it (deterministic character of the protocol). 
The alphabet $A$ can be identified with the Galois field $G$.
Moreover, let us consider the unitary transformations $V_0^a$ for 
$a \in A$, defined by
	\beq
V^a_0\, \ket{v^0_t} 
= \omega^{t \odot a} \ket{v^0_t}, 
	\eeq
which can be regarded as the generalized Pauli $Z$ operators. 

Then, Alice encoding operation will be the shift operation realized by the 
operator $V^a_0$ with $a\in A$ on all the MUB but the computational one, 
that is for $k > 0$: 
	\beq
V^a_0\, \ket{v^k_t}
= {1 \over \sqrt d} \sum_{q=0}^{d-1}
\omega^{\ominus q \odot (t \ominus a)}
(\omega^{(k-1) \odot q \odot q})^{\frac{1}{2}} \ket{v^0_q}
= \ket{v^k_{t \ominus a}}.
	\eeq

In such a case, Bob receiving back the state $\ket{v^k_{t \ominus a}}$ can 
unambiguously determine $a$ by means of a projective measurement onto the $k$-th 
basis. In fact, he will get the value 
	\beq \label{bval}
b = t \ominus a
	\eeq
from which, knowing $t$, he can extract $a$.

\medskip

Then, the protocol runs as follows:
\begin{itemize}
	\item[1.] 
	Bob randomly prepares one of the $d^2$ qudit states $\ket{v_{t}^{k}}$,
	with $k = 1, \ldots, d$ and 
	$t = 0, \ldots, d-1$, and sends it to Alice.
	\item[2.] 
	Alice, upon receiving the qudit state has two options.
		\begin{itemize}
			\item[a)] 
			  With probability $c \neq 0$, she performs a measurement 
			by projecting over a randomly chosen basis among the $d$ bases 
				with $k = 1, \ldots, d$ (\textit{Control Mode}). 
			 She then sends back to Bob the resulting state. 
			\item[b)] 
			  With probability $1- c$, she encodes a symbol 
			  $a \in A$ by applying the unitary operator
			  $V_0^{a}$ (\textit{Message Mode}).
			She then sends back to Bob the resulting state. 
		\end{itemize}
	\item[3.] 
	Bob, upon receiving back the qudit state, performs a measurement by 
	projecting over the basis to which the qudit state initially belonged.
	\item[4.] 
	At the end of the transmission, Alice publicly declares on which runs she 
	performed the control mode and on which others the message mode.
	In the first case, Alice announces the bases over which she measured.
	Then, by public discussion, a comparison of Alice's and Bob's 
	measurements results is performed over coincident bases.
	In the ideal case (noiseless channels and no eavesdropping) their results 
	must coincide.
	
	In the message mode runs, Bob gets the encoded symbol $a$ as discussed above.
	
\end{itemize}
Notice at the above point 2. the deterministic character of the protocol given 
by the possibility for Alice, besides to decide when to encode, to determine 
the message (key) sequence, since she knows that Bob will unambigously decode 
each character of the message (key).


\section{Security of the protocol} 

Among individual attacks the most elementary one is the \emph{Intercept-Resend}.
Suppose Eve, to learn Alice's operation, performs projective measurements on 
both paths of the traveling qudit, randomly choosing the measuring basis. 
She will steal the whole information for each message mode run, indipedently 
from the chosen basis. 
However, in each control mode run with coincident bases for Alice and Bob, she can 
guess the correct basis with probability $1/d$, and in this case she is not detected 
at all. 
If otherwise Eve chooses the wrong basis, she still has a probability $1/d$ to evade 
detection on the forward path and probability $1/d$ on the backward path, leading to 
an overall probability $1/d^{2}$ to remain undetected. 
This means that the double test of Alice and Bob reveals Eve with probability 
$(d^2-1)(d-1)/d^4$, including the cases of non-coincident bases.

\begin{figure}[htbp] 
\begin{center}
\includegraphics[scale=1]{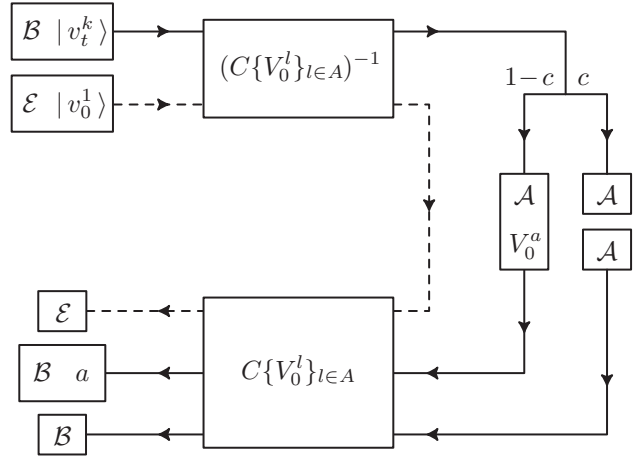}
\end{center}\smallskip
\caption{The scheme summarizing our protocol. 
Labels $\mathcal{B}$ and $\mathcal{E}$ stand for Bob's and Eve's qudit systems 
respectively. Label $\mathcal{A}$ denotes Alice's operation on Bob's qudit. 
$(C\{V^l_0\}_{l\in A})^{-1}$ and $C\{V^l_0\}_{l\in A}$ represent the eavesdropping 
operations on the forward and backward path respectively.}
 \label{protocol_scheme} 
	\end{figure}

We are going to prove the security of the protocol against a more powerful 
individual attack. Quite generally, in individual attacks Eve lets the carrier of 
information interact with an ancilla system she has prepared and then try to gain 
information by measuring the ancilla. 
In this protocol she has to do that two times,  in the forward path 
(to gain information about the state Bob sends to Alice) and in the backward path 
(to gain information about the state Alice sends back to Bob, hence about Alice's 
transformation).
Moreover, by using the same ancilla in the forward and backward path, Eve could 
benefit from quantum interference effects (see Fig.~\ref{protocol_scheme}).

In particular, we consider the unitary transformation describing the attack 
as controlled shifts $\{V^l_0\}_{l\in A}$, where the controller is the 
traveling qudit, while the target is in the Eve's hands. 
That is, $C\{V^l_0\}_{l\in A}:\mathcal{H}_{d}\otimes\mathcal{H}_{d}
\to\mathcal{H}_{d}\otimes\mathcal{H}_{d}$ defined as follows:  
	\beq \label{CV^a_0}
\ket{v^1_{t_1}} \ket{v^1_{t_2}}\longarrow{C\{V^l_0\}_{l\in A}}
\ket{v^1_{t_1}} V^{l=t_1}_0 \ket{v^1_{t_2}}
= \ket{v^1_{t_1}} \ket{v^1_{{t_2} \ominus {t_1}}}.
	\eeq
We remark that, in this definition, the controller as well as the target 
states are considered in the dual basis for the sake of simplicity. 
Other choices (except the computational basis) will give the same final 
results.

Then, we consider Eve intervening in the forward path with 
$(C\{V^l_0\}_{l\in A})^{-1}$, defined by 
	\beq
\ket{v^1_{t_1}}\ket{v^1_{t_2}}\longarrow{(C\{V^l_0\}_{l\in A})^{-1}}
\ket{v^1_{t_1}} V^{\ominus {t_1}}_0 \ket{v^1_{t_2}}
= \ket{v^1_{t_1}} \ket{v^1_{{t_2} \ominus ({\ominus t_1})}}
= \ket{v^1_{t_1}} \ket{v^1_{{t_2} \oplus {t_1}}},
	\eeq
and with $C\{V^l_0\}_{l\in A}$ in the backward path. 

\medskip
%


\subsection{Message Mode}

Now, let us analyze in detail the transformations of the quantum states 
on an entire message mode run.

\medskip

\noindent {\it Attack on the forward path.}

\smallskip

The initial Bob state is one of the $d^{2}$ states $\ket{v_{t}^{k}}$, with 
$k = 1, \ldots, d$ and $t = 0, \ldots, d-1$. 
Then, Eve initially prepares the ancilla state $\ket{v_{0}^{1}}_{\mathcal E}$ 
in the dual basis and performs the controlled operation. Hence, we get
	\beq \label{1attack}
\ket{v^k_t}_{\mathcal B} \ket{v^1_0}_{\mathcal E} 
\longarrow{(C\{V^l_0\}_{l\in A})^{-1}}
\sum_{h=0}^{d-1}
\bracket{v^1_h}{v^k_t}
\ket{v^1_h}_{\mathcal B} \ket{v^1_0}_{\mathcal E} =
\sum_{h=0}^{d-1}
\bracket{v^1_h}{v^k_t}
\ket{v^1_h}_{\mathcal B} \ket{v^1_h}_{\mathcal E}.
	\eeq

\medskip

\newpage

\noindent {\it Encoding.}

\smallskip

The Bob's qudit state undergoes the shift $V_0^a$ with $a \in A$, 
then from (\ref{1attack}) we get
	\beq \label{cod}
\arrow{V^a_0} \ \sum_{h=0}^{d-1}
\bracket{v^1_h}{v^k_t}
\ket{v^1_{h \ominus a}}_{\mathcal B}
\ket{v^1_h}_{\mathcal E}.
	\eeq

\medskip

\noindent {\it Attack on the backward path.}

\smallskip

The state (\ref{cod}) undergoes a $C\{V^l_0\}_{l\in A}$ operation, 
hence we have
	\beq \label{2attack}
\longarrow{C\{V^l_0\}_{l\in A}} \ \sum_{h=0}^{d-1}
\bracket{v^1_h}{v^k_t}
\ket{v^1_{h \ominus a}}_{\mathcal B}
\ket{v^1_{h \ominus (h \ominus a)}}_{\mathcal E}
= \sum_{h=0}^{d-1}
\bracket{v^1_h}{v^k_t}
\ket{v^1_{h \ominus a}}_{\mathcal B} \ket{v^1_a}_{\mathcal E} 
= \ket{v^k_{t \ominus a}}_{\mathcal B} \ket{v^1_a}_{\mathcal E}.
	\eeq

Then, Eve measures her ancilla system by projecting in the dual basis, according to 
the chosen initial ancilla state.

\medskip

We notice that the controlled operations performed by Eve, as well as her final 
measurement, left unchanged Bob's qudit state. 
Hence, Bob's measurement by projection in the $k$-th basis to which the initial 
state belonged, always allows him to obtain the symbol $a$ Alice has encoded 
[see (\ref{bval})].

On the other hand, Eve gets $\ket{v^1_a}$ with probability 1 as the result of her 
measurement. Therefore, she is able to exactly determine the encoded symbol $a$ 
as well and she steals the whole information, quantified in bits, 
	\beq \label{I_E}
I_\mathcal{E} = \log_2{d}
	\eeq
on each message mode run.


\subsection{Control Mode}

We would like to evaluate the probability $P_{\mathcal E}$ Alice and Bob have 
to reveal Eve on each control mode run. Alice and Bob only compare the results 
of their measurements when, by public discussion, they agree on the used basis. 

Let us focus on the case Alice and Bob use the same basis $k$, keeping in mind 
that it happens with probability $1/d$.
The situation is different for $k=1$ and $k \neq 1$, due to the Eve's choice
of using the dual basis for her ancilla. 
\begin{itemize}

\item[1)] 
For $k=1$, on the forward path we have
	\beq
\ket{v_t^{1}}_{\mathcal B}\ket{v^1_0}_{\mathcal E} 
\longarrow{(C\{V^l_0\}_{l\in A})^{-1}}
\ket{v^1_t}_{\mathcal B} \ket{v^1_t}_{\mathcal E}.
	\eeq
Alice, measuring in the dual basis, gets ${\bar t}$ with probability 1 and 
projects into $\ket{\bar t}_{\mathcal B}$. On the backward path we have
	\beq
\ket{v^1_t}_{\mathcal B} \ket{v^1_t}_{\mathcal E}
\; \longarrow{C\{V^l_0\}_{l\in A}} \;
\ket{v^1_t}_{\mathcal B} \ket{v^1_{t \ominus t}}_{\mathcal E}
= \ket{v^1_t}_{\mathcal B} \ket{v^1_0}_{\mathcal E}.
	\eeq
Bob, in turn, by measuring in the dual basis gets $t$ with probability 1.
Thus, Alice and Bob have perfect correlation and $P_{\mathcal E}=0$. 
\item[2)] 
For $k = 2, \ldots,  d$, we get on the forward path
	\beq \label{CM_2)1}
\ket{v^k_t}_{\mathcal B} \ket{v^1_0}_{\mathcal E}
= \sum_{h=0}^{d-1}
\bracket{v^1_h}{v^k_t}
\ket{v^1_h}_{\mathcal B} \ket{v^1_0}_{\mathcal E}
\longarrow{(C\{V^l_0\}_{l\in A})^{-1}}
\sum_{h=0}^{d-1}
\bracket{v^1_h}{v^k_t}
\ket{v^1_h}_{\mathcal B} \ket{v^1_h}_{\mathcal E}.
	\eeq
By expressing the vectors of the dual basis in terms of the basis $k$ used by Bob, 
we rewrite the right hand side of (\ref{CM_2)1}) as
	\beq \label{CM_2)k}
\sum_{h=0}^{d-1}
\bracket{v^1_h}{v^k_t}
\sum_{s=0}^{d-1}
\bracket{v^k_s}{v^1_h}
\ket{v^k_s}_{\mathcal B} \ket{v^1_h}_{\mathcal E}.
	\eeq
At this point Alice measures in the basis $k$.
The result of her measurement is to project into $\ket{v^k_{t'}}$, 
whatever $t' \in A$ is, with probability   
	\beq  
\sum_{h=0}^{d-1}
|\bracket{v^1_h}{v^k_t}
\bracket{v^k_{t'}}{v^1_h}|^2 = 
\sum_{h=0}^{d-1}
|\bracket{v^1_h}{v^k_t}|^2
|\bracket{v^k_{t'}}{v^1_h}|^2 = 
\sum_{h=0}^{d-1} {1 \over d^2} = {1 \over d}
	\eeq
according to definition of MUB.

Among the $d$ possibilities we distinguish two cases. 
\begin{itemize}

\item [a)] $t' = t$, occurring with probability $1/d$, for which the resulting 
state from (\ref{CM_2)k}) is	
	\beq \label{result_a} 
\sqrt d
\sum_{h=0}^{d-1}
\bracket{v^1_h}{v^k_t}
\bracket{v^k_t}{v^1_h}
\ket{v^k_t}_{\mathcal B} \ket{v^1_h}_{\mathcal E}
= \frac{1}{\sqrt d}
\sum_{h=0}^{d-1}
\ket{v^k_t}_{\mathcal B} \ket{v^1_h}_{\mathcal E}.
	\eeq
We have now to apply the $C\{V^l_0\}_{l\in A}$ operation of the 
backward path. Thus, (\ref{result_a}) transforms as follows
	\beq
\ket{v^k_t}_{\mathcal B}
{1 \over {\sqrt d}}
\sum_{h=0}^{d-1} \ket{v^1_h}_{\mathcal E} 	
= \sum_{h'=0}^{d-1}
\bracket{v^1_{h'}}{v^k_t}
\ket{v^1_{h'}}_{\mathcal B} {1 \over {\sqrt d}}
\sum_{h=0}^{d-1} \ket{v^1_h}_{\mathcal E}
	\eeq
	\beq
\longarrow{C\{V^l_0\}_{l\in A}}
\sum_{h'=0}^{d-1}
\bracket{v^1_{h'}}{v^k_t}
\ket{v^1_{h'}}_{\mathcal B} {1 \over {\sqrt d}}
\sum_{h=0}^{d-1} \ket{v^1_{h \ominus h'}}_{\mathcal E} = 
\ket{v^k_t}_{\mathcal B} 
{1 \over {\sqrt d}} \sum_{r=0}^{d-1} \ket{v^1_r}_{\mathcal E},
	\eeq
where $r = h \ominus h'$.

It results that Eve's attack does not alter the eigenvector 
$\ket{v_t^k}_{\mathcal B}$.
Hence, Bob upon his measurement will get $t$ with probability 1.
Then, neither Alice nor Bob outwit Eve's attacks.

\item [b)] $t' \neq t$, occurring with probability $(d-1)/d$, for which Alice, 
getting a state different from the one initially sent by Bob, outwits Eve 
in the forward path. Hence, in this case, we do not need to explicitly evaluate 
the state change in the backward path.

\end{itemize}
\end{itemize}

In summary, from the analyzed cases, we have:
\begin{itemize}

\item $1/d$ 
the probability with which Bob and Alice measure in the same basis $k$;

\item $(d-1)/d$  
the probability of Bob choosing the initial state $\ket{v_t^k}$ from any basis
but the dual one, that is $k \neq 1$;
     
\item $(d-1)/d$ 
the probability that the state $\ket{v_t^k}$ sent by Bob gives a measurement 
result $\ket{v_{t'}^k}$ with $t' \neq t$ to Alice.

\end{itemize}

We then conclude that the probability for Alice and Bob to outwit Eve on each 
control mode run is
	\beq \label{P_E}
P_{\mathcal E}=
\frac{1}{d} \cdot \frac{d-1}{d} \cdot \frac{d-1}{d}=
{\frac{(d-1)^{2}}{d^{3}}} \, .
	\eeq

\begin{figure}[htbp]
\begin{center}
\includegraphics[scale=1]{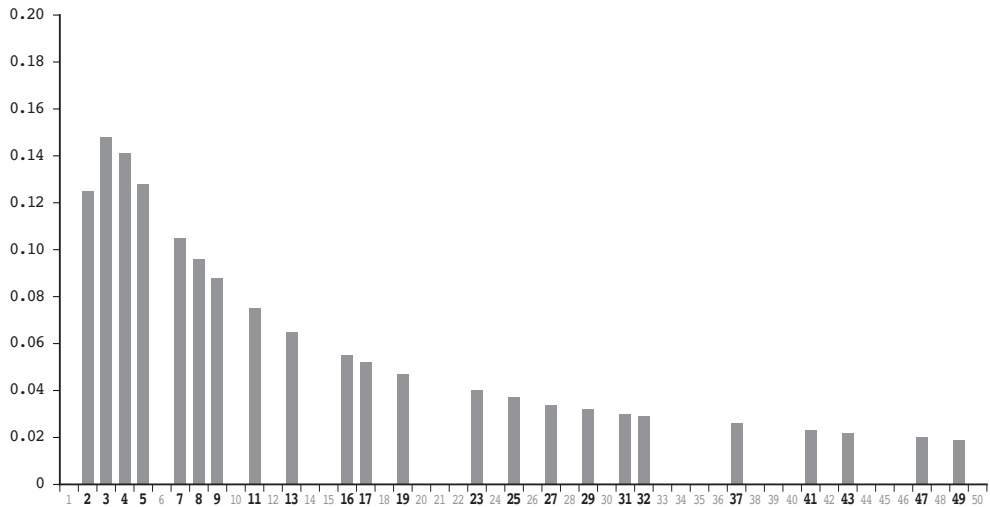}
\end{center}\smallskip
\caption{The probability $P_{\mathcal E}$ versus the dimension $d$ 
(bars correspond to prime power numbers).} 
\label{graf_Pe(new)}
	\end{figure}

In Fig.~\ref{graf_Pe(new)} we show the behavior of $P_{\mathcal E}$ versus the 
order $d$ of the alphabet. 
Interestingly enough, the values of $P_{\mathcal E}$ at $d=3,4,5$ are higher than 
that at $d=2$.
In particular, $P_{\mathcal E}$ has a maximum at $d=3$ showing that this dimension 
represents the optimal compromise between two different trends. 
On the one hand, the probability $(d-1)^2/d^2$ of revealing Eve in each successful 
control mode run (that is when the bases of Alice and Bob coincide) increases 
towards 1 when increasing the dimension $d$. 
On the other hand, the efficiency of the whole control process decreases according 
to the probability $1/d$ for each control mode run to succeed.
\medskip
%


\section{Concluding remarks}

We have proposed a deterministic cryptographic protocol working with a $d$-ary 
alphabet and exploiting a bidirectional quantum channel.
When considering an attack performed by means of controlled operations on both 
directions of the quantum channel, we have found that Eve can steal the total 
amount of information $I_{\mathcal E}$ (see (\ref{I_E})), while the probability 
$P_{\mathcal E}$ to outwit her presents a maximum for $d=3$ (see (\ref{P_E})).

Contrarily to probabilistic protocols, the deterministic nature of this protocol 
also allows the realization of Quantum Direct Communication (QDC) between legitimate 
users \cite{BEKW02, BF02, CL04, LM05}. 
In this case Alice and Bob (after authentication) can communicate directly the 
meaningful message without encryption. 
However, for this kind of communication only an asymptotic security can be proven. 
In fact, if we assume that Eve wants to perform her attack on each message mode run, 
without having been detected in the previous control mode runs, then the probability 
is given by following geometric series:
\beq
 (1-c) + c(1-P_{\mathcal E})(1-c) + 
c^{2}(1-P_{\mathcal E})^{2}(1-c) + \ldots 
=\frac{1-c}{1-c(1-P_{\mathcal E})}.
\eeq

Thus, being $I_{\mathcal E}$ the quantity of information that Eve eavesdrops in 
a single attack, the probability that she successfully eavesdrops an amount of 
information $I$ is
\beq \label{fmla_QDC}
\left(\frac{1-c}{1-c\big(1-P_{\mathcal E}\big)}\right)
^{\!I/I_{\mathcal E}},
\eeq
with $I_{\mathcal E}$ and $P_{\mathcal E}$ given in (\ref{I_E}) and (\ref{P_E}) 
respectively. 

We observe that such a probability exponentially decreases towards 0 as a function 
of $I$ for each given dimension $d$. 
So, (\ref{fmla_QDC}) expresses the asymptotic security of the direct communication 
use of the protocol.

However, in this case the probability for Alice and Bob to detect Eve before 
she can eavesdrop a fixed amount of information, that is the complement of 
probability in (\ref{fmla_QDC}), is maximal for $d=2$. 

It is interesting to notice that the optimal dimension depends on the specific 
task of the protocol (QKD or QDC). Therefore, we believe that this work might 
open up new horizons for deterministic cryptographic protocols involving finite 
dimensional systems.



\appendix

\section{}

By referring to \cite{D05}, let us denote by $V^{j}_{l}$ the operators
given by the composition of the shifts in the computational and the dual 
basis, that is 
	\beq
V^{j}_{l} = V^{j}_{0} \cdot V^{0}_{l} = \sum_{t=0}^{d-1}\omega^{(t\oplus l)\odot j}
\ket{t \oplus l}\bra{t}.
	\eeq

This set of operators coincides with the Generalized Pauli Group 
(see \cite{BBRV01}).
The $V^{j}_{l}$'s are $d^2$ unitary transformations which satify the following 
composition law
	\beq \label{prV}
V^{j}_{l} \cdot V^{j'}_{l'} = \omega^{(l \odot j')} V^{j \oplus j'}_{l \oplus l'},
	\eeq
and, up to phases, they form $d+1$ commuting subgroups of $d$ elements that have only 
the identity in common.
The $k$-th subgroup, with $k=0, \ldots, d$, admits $\{\ket{v_t^k}\}_{t=0,\ldots, d-1}$ 
as diagonalizing basis. 
Its elements are denoted by $U^{k}_{l}$ with $l=0, \ldots, d-1$, 
and they are required to satisfy:
	\beq \label{prU}
U^{k}_{l \oplus l'} = U^{k}_{l} \cdot U^{k}_{l'} \, ,
	\eeq
	\beq \label{diag}
U^{k}_{l} = \sum_{t=0}^{d-1}\omega^{t\odot l}\ket{v_t^k}\bra{v_t^k} \, , 
	\eeq
	\beq \label{ph}
U^{k}_{l} = V^{(k-1)\odot l}_{l} 
\quad \text{up to a phase which is 1 for $l=0$}.
	\eeq
It is important to point out that (\ref{prV}), (\ref{prU}), 
(\ref{diag}) and (\ref{ph}) must be guaranteed at the same time.
In \cite{D05}, the following relation is obtained from them:
	\beq \label{DetSgn}
U^{k}_{l} = 
(\omega^{\ominus (k-1)\odot l \odot l})^{\frac{1}{2}} \, V^{(k-1)\odot l}_{l}
	\eeq

In the odd prime power case such expression is completely determined and the phase 
is a $p$-th root of unity. In fact the square root can be interpreted as the division 
of the exponent by 2 in the Galois field $G$. 

This is no longer true in the even prime power case. In this case the phase is not a 
$2$-nd root of unity but a $4$-th root of unity, that is it can also assume the 
values $\pm i$, other than $\pm 1$. 
Moreover, the sign of it is still undetermined. 
The determination of such sign provided in \cite{D05} is uncorrect.

Below we correctly develop the last step of (32) in \cite{D05} getting the right 
sign, and consequently the square root's sign in (\ref{ket_d}), as indicated 
in (\ref{srs}).  

First of all, we observe that for $p=2$ we have $\omega=-1$.

In \cite{D05} it has been implicitly chosen the determination of the square root 
of $\omega^{(k-1) \odot 2^n \odot 2^n}$ as to be $i^{(k-1)\odot 2^n \odot 2^n}$.
Then, we have:
	\beqa
U^k_l & \!\!\! = \!\!\! &
\prod^{m-1}_{n=0} U^k_{l_n \odot 2^n} = 
\prod^{m-1}_{n=0} {(U^k_{2^n})}^{l_n} \nonumber \\
& \!\!\! = \!\!\! & \prod^{m-1}_{\textstyle{n=0 \atop l_n \neq 0}} 
\! {(\omega^{(k-1) \odot 2^n \odot 2^n})}^{\frac{1}{2}}
(V^{(k-1) \odot 2^n}_{2^n}) \nonumber \\
& \!\!\! = \!\!\! & \left( 
\prod^{m-1}_{\vrule height9pt width0pt 
\smash{\textstyle{n=0 \atop l_n \neq 0}}} 
\! i^{(k-1) \odot 2^n \odot 2^n}
\! \right) \! \!
\left( 
\prod^{m-1}_{\vrule height9pt width0pt 
\smash{\textstyle{n=0 \atop l_n \neq 0}}} 
\! V^{(k-1) \odot 2^n}_{2^n}
\! \right).
	\eeqa

\newpage
Let $n_0, n_1, \ldots n_h$ be the indices $n_j$ such that $l_{n_j}=1$. 
By taking into account (\ref{prV}), the second product can be rewritten 
as follows.

	\beqa
\prod^{m-1}_{\textstyle{n=0 \atop l_n \neq 0}} (V^{(k-1) \odot 2^n}_{2^n}) 
& \!\!\! = \!\!\! & \prod^{h}_{j=0} V^{(k-1) \odot 2^{n_j}}_{2^{n_j}} \nonumber \\[-6pt]
& \!\!\! = \!\!\! & \left( 
\prod^{h}_{j=1} \omega^{(k-1) \odot 
(2^{n_0} \oplus 2^{n_1} \oplus \ldots \oplus 2^{n_{j-1}}) \odot 2^{n_j}}
\! \right)
V^{(k-1) \odot (2^{n_0} \oplus \ldots \oplus 2^{n_{h}})}
_{(2^{n_0} \oplus \ldots \oplus 2^{n_{h}})} \nonumber \\[2pt]
& \!\!\! = \!\!\! & \left(
\prod^{h}_{j=0} \omega^{(k-1) \odot 2^{n_j} \odot (l \mod 2^{n_j})} 
\! \right)
V^{(k-1) \odot l}_{l} \nonumber \\[2pt]
& \!\!\! = \!\!\! & \left(
\prod^{m-1}_{\vrule height9pt width0pt \smash{\textstyle{n=0 \atop l_n \neq 0}}}
\omega^{(k-1) \odot 2^{n} \odot (l \mod 2^{n})} 
\! \right)
V^{(k-1) \odot l}_{l}.
	\eeqa

Then, we have:
	\beq
U^k_l =
\left(
\prod^{m-1}_{\vrule height9pt width0pt \smash{\textstyle{n=0 \atop l_n \neq 0}}} 
\! i^{(k-1) \odot 2^n \odot 2^n}
\omega^{(k-1) \odot 2^{n} \odot (l \mod 2^{n})}
\! \right)
V^{(k-1) \odot l}_{l}.
	\eeq

This gives the correct determination of square root's sign in the phase 
as in (\ref{srs}), which can be rewritten as
\beq \label{srs2}
\prod_{\vrule height9pt width0pt \smash{\textstyle{n=0 \atop l_n \neq 0}}}^{m-1} \!
i^{(k-1)\odot 2^n \odot 2^n} 
\omega^{(k-1) \odot 2^n \odot (l \mod 2^n)}= 
\prod_{n=0}^{m-1}
(-1)^{\sum_{h=0}^{n-1}l_n l_h (k-1) \odot 2^n \odot 2^h} 
i^{l_n(k-1)\odot 2^n \odot 2^n}. 
\eeq

\medskip

Now, by referring to (\ref{OperExp_G}), we remark that an analogous property 
does not hold for powers of $i$ with exponents in $G$.
The reader can easily check that
	\beq \label{*}
i^j \cdot i^l 
= (-1)^{j l} i^{j \oplus l}
= (-1)^ {j_0 l_0} i^{j \oplus l}.
	\eeq

From ($\ref{*}$) it follows that
	\beq \label{**}
(\omega^{(k-1) \odot l \odot l})^{1/2}
(\omega^{(k'-1) \odot l \odot l})^{1/2} =
\phi(k,k',l) \, (\omega^{((k-1) \oplus (k'-1)) \odot l \odot l})^{\frac{1}{2}},
	\eeq
where we have defined
	\beq
\phi(k,k',l) = (-1)^{\sum_{n=0}^{m-1}l_n((k-1)\odot 2^n 
\odot 2^n)((k'-1)\odot 2^n \odot 2^n)}.
	\eeq

\medskip 

In fact, by using (\ref{srs2}) and (\ref{*}),
	\beqa
& & \kern-75pt (\omega^{(k-1) \odot l \odot l})^{\frac{1}{2}}
(\omega^{(k'-1) \odot l \odot l})^{\frac{1}{2}} \nonumber \\[2pt]
& & \kern-75pt \kern20pt {} = \, \prod_{n=0}^{m-1} 
(-1)^{\sum_{h=0}^{n-1}l_nl_h (k-1) \odot 2^n \odot 2^h}
i^{l_n(k-1)\odot 2^n \odot 2^n} \nonumber \\[-6pt]
& & \kern-75pt \kern60pt {} \times (-1)^{\sum_{h=0}^{n-1}l_nl_h (k'-1) \odot 2^n \odot 2^h}
i^{l_n(k'-1)\odot 2^n \odot 2^n} \nonumber
	\eeqa
\vspace{-20pt}
	\beqa
& & \kern20pt {} = \, \prod_{n=0}^{m-1}
(-1)^{\sum_{h=0}^{n-1}l_n l_h ((k-1)\oplus(k'-1)) \odot 2^n \odot 2^h}
\nonumber \\[-6pt]
& & \kern60pt {} \times (-1)^{(l_n(k-1)\odot 2^n \odot 2^n)(l_n(k'-1)\odot 2^n \odot 2^n)}
\, i^{l_n((k-1) \oplus(k'-1))\odot 2^n \odot 2^n} \nonumber \\[8pt]
& & \kern20pt {} = \, \phi(k,k',l) \, 
(\omega^{((k-1) \oplus (k'-1)) \odot l \odot l})^{\frac{1}{2}}.
	\eeqa

\medskip

\noindent
In particular, by assuming $k' = k$ in (\ref{**}), we get the conjugate of 
$(\omega^{(k-1) \odot q \odot q})^{\frac{1}{2}}$ as
	\beq
\phi(k,k,q) \, (\omega^{(k-1) \odot q \odot q})^{\frac{1}{2}}.
	\eeq

Consequently, the correct expression for the inner products 
$\bracket{v_{t'}^{k'}}{v_t^k}$ with $k,k' \geq 1$ is the following 
(which does not coincide with (28) in \cite{D05}):
	\beqa \label{InnerProd}
\bracket{v_{t'}^{k'}}{v_t^k} & \!\!\! = \!\!\! &
{1 \over d} \sum_{q=0}^{d-1} 
\omega^{q \odot t}
(\omega^{(k-1) \odot q \odot q})^{\frac{1}{2}}
\, \omega^{q \odot t'}
\phi(k',k',q) \,
(\omega^{(k'-1) \odot q \odot q})^{\frac{1}{2}}
\nonumber \\
& \!\!\! = \!\!\! & 
{1 \over d} \sum_{q=0}^{d-1}
\phi(k,k',q)
\, \phi(k',k',q) \,
\omega^{q \odot (t \oplus t')}
(\omega^{((k-1)\oplus (k'-1)) \odot q \odot q})^{\frac{1}{2}}.
	\eeqa

In order to prove the MUB condition, we state the following elementary 
properties of the function $\phi$:
	\beqa
\phi(k,k',0) & \!\!\! = \!\!\! & 1 \label{prop1}\\
\phi(k',k,q) & \!\!\! = \!\!\! & \phi(k,k',q) \label{prop2}\\ 
\phi(k,k',q) \, \phi(k,k',q') & \!\!\! = \!\!\! & \phi(k,k',q\oplus q') \label{prop3}\\ 
\phi(k,k,q) & \!\!\! = \!\!\! & \omega^{(k-1) \odot q \odot q} \label{prop4}
	\eeqa
The first and the second one come from the very definition of $\phi$, 
the third one comes from the fact that $q_n + q'_n \mod 2 = (q \oplus q')_n$ and 
the fourth one from (\ref{**}) for $k'=k$.

We also need to verify that the following equality, corresponding to (37) 
in \cite{D05},
	\beq \label{37Durt}
(\omega^{(k-1) \odot q \odot q})^{\frac{1}{2}}
(\omega^{(k-1) \odot q' \odot q'})^{\frac{1}{2}}
=
\omega^{(k-1) \odot q \odot q'}
(\omega^{(k-1) \odot (q \oplus q') \odot (q \oplus q')})^{\frac{1}{2}}
	\eeq
holds with the correct determination of square root's sign given by (\ref{srs2}) 
(this does not happen with wrong determination of the sign given in \cite{D05}).

Let us consider the left hand side. It turns out to be
	\beqa
& & \kern-10pt (\omega^{(k-1) \odot q \odot q})^{\frac{1}{2}}
(\omega^{(k-1) \odot q' \odot q'})^{\frac{1}{2}} \nonumber \\
& & \kern-5pt {} = \, \prod_{n=0}^{m-1} 
(-1)^{\sum_{h=0}^{n-1} q_n q_h (k-1) \odot 2^n \odot 2^h}
(-1)^{\sum_{h=0}^{n-1}q'_n q'_h (k-1) \odot 2^n \odot 2^h}
i^{q_n (k-1) \odot 2^n \odot 2^n}
i^{q'_n (k-1) \odot 2^n \odot 2^n} \nonumber \\
& & \kern-5pt {} = \, \prod_{n=0}^{m-1}
(-1)^{\sum_{h=0}^{n-1} (q_n q_h + q'_n q'_h ) (k-1) \odot 2^n \odot 2^h}
(-1)^{q_n q'_n (k-1) \odot 2^n \odot 2^n}
i^{(q \oplus q')_n (k-1) \odot 2^n \odot 2^n} \nonumber \\[4pt]
& & \kern-5pt {} = \, (-1)^{
\left(
\sum_{n=0}^{m-1}q_n q'_n (k-1) \odot 2^n \odot 2^n
\right)
+ 
\left(
\sum_{n=0}^{m-1}\sum_{h=0}^{n-1} 
(q_n q_h + q'_n q'_h ) (k-1) \odot 2^n \odot 2^h
\right)} \nonumber \\[-4pt]
& & \kern-5pt \kern220pt {} \times \prod_{n=0}^{m-1}
\! i^{(q \oplus q')_n (k-1) \odot 2^n \odot 2^n}. 
	\eeqa

For the right hand side, we have:
	\beqa
& & \omega^{(k-1) \odot q \odot q'}
(\omega^{(k-1) \odot (q \oplus q') \odot (q \oplus q')})^{\frac{1}{2}}
\nonumber \\[4pt]
& & \kern20pt {} = \, (-1)^{(k-1) \odot (\sum_{n=0}^{m-1} q_n 2^n) \odot (\sum_{n=0}^{m-1} q'_n 2^n)}
\nonumber \\[-2pt]
& & \kern60pt {} \times \prod_{n=0}^{m-1}
(-1)^{\sum_{h=0}^{n-1}(q \oplus q')_n (q \oplus q')_h (k-1) \odot 2^n \odot 2^h} 
i^{(q \oplus q')_n (k-1)\odot 2^n \odot 2^n} \nonumber \\[4pt]
& & \kern20pt {} = \, (-1)^{
\left(\sum_{n=0}^{m-1}\sum_{h=0}^{m-1}
q_n q'_h (k-1) \odot 2^n \odot 2^h
\right)
+ 
\left(\sum_{n=0}^{m-1}\sum_{h=0}^{n-1}
  (q \oplus q')_n (q \oplus q')_h (k-1) \odot 2^n \odot 2^h
\right)} \nonumber \\[-2pt]
& & \kern60pt {} \times \prod_{n=0}^{m-1}
i^{(q \oplus q')_n (k-1)\odot 2^n \odot 2^n}.
	\eeqa

At this point, (\ref{37Durt}) derives from the following sequence of 
equalities mod $2$:
	\beqa
& & \kern-20pt \left(
\sum_{n=0}^{m-1}\sum_{h=0}^{m-1}
q_n q'_h 2^n \odot 2^h
\! \right)
+
\left(
\sum_{n=0}^{m-1}\sum_{h=0}^{n-1}
(q \oplus q')_n (q \oplus q')_h 2^n \odot 2^h
\! \right) \nonumber \\
& & {} = \, \left(
\sum_{n=0}^{m-1}\sum_{h=0}^{m-1}
q_n q'_h 2^n \odot 2^h
\! \right)
+
\left(
\sum_{n=0}^{m-1}\sum_{h=0}^{n-1}
(q_n + q'_n) (q_h + q'_h) 2^n \odot 2^h
\! \right) \nonumber \\
& & {} = \, \left(
\sum_{n=0}^{m-1}
q_n q'_n 2^n \odot 2^n
\! \right)
+
\left(
\sum_{n=0}^{m-1}\sum_{h=0}^{n-1}
(q_n q_h + q'_n q'_h) 2^n \odot 2^h
\! \right).
	\eeqa

Finally, we can prove the MUB condition for even prime power.

\smallskip

From (\ref{InnerProd}), by using in the order (\ref{prop2}), (\ref{prop3}), 
(\ref{37Durt}) and (\ref{prop4}), and then relabelling the sum indices, we have
	\beqa
& & \kern-10pt \bracket{v_{t'}^{k'}}{v_t^k} \bracket{v_t^k}{v_{t'}^{k'}} 
\nonumber \\[4pt]
& & \kern-10pt \kern20pt {} = \, {1 \over d^2}
\sum_{q,q'=0}^{d-1}
\phi(k,k',q) \, \phi(k',k',q)
\, \phi(k,k',q') \, \phi(k,k,q') \nonumber \\[-4pt]
& & \kern-10pt \kern80pt {} \times \omega^{q \odot (t \oplus t')}
\omega^{q' \odot (t \oplus t')}
(\omega^{((k-1)\oplus (k'-1)) \odot q \odot q})^{\frac{1}{2}}
(\omega^{((k-1) \oplus (k'-1)) \odot q' \odot q'})^{\frac{1}{2}} 
\nonumber \\
& & \kern-10pt \kern20pt {} = \, {1 \over d^2}
\sum_{q,q'=0}^{d-1}
\phi(k,k',q \oplus q')
\, \phi(k,k,q \oplus q')
\, \phi(k,k,q)
\, \phi(k',k',q) \nonumber \\[-4pt]
& & \kern-10pt \kern80pt {} \times
\omega^{(q \oplus q') \odot (t \oplus t')}
\omega^{((k-1) \oplus (k'-1)) \odot q \odot q'}
(\omega^{((k-1)\oplus (k'-1)) \odot (q \oplus q') 
\odot (q \oplus q')})^{\frac{1}{2}} \nonumber \\
& & \kern-10pt \kern20pt {} = \, {1 \over d^2}
\sum_{q,h=0}^{d-1}
\phi(k,k',h)
\, \phi(k,k,h)
\, \omega^{(k-1) \odot q \odot q}
\omega^{(k'-1) \odot q \odot q} \nonumber \\[-4pt]
& & \kern-10pt \kern80pt {} \times
\omega^{h \odot (t \oplus t')}
\omega^{((k-1) \oplus (k'-1)) \odot q \odot (q \oplus h)}
(\omega^{((k-1)\oplus (k'-1)) \odot h \odot h})^{\frac{1}{2}}.
\nonumber	
	\eeqa
\noindent Now, by collecting the terms without $q$ and then using 
(\ref{la4}), the previous expression can be rewritten as
	\beqa
& & \kern-20pt {1 \over d^2}
\sum_{h=0}^{d-1}
\phi(k,k',h)
\, \phi(k,k,h)
\, \omega^{h \odot (t \oplus t')}
(\omega^{((k-1)\oplus (k'-1)) \odot h \odot h})^{\frac{1}{2}}
\sum_{q=0}^{d-1}
\omega^{((k-1) \oplus (k'-1)) \odot q \odot h} 
\nonumber 
	\eeqa
\vspace{-10pt}
	\beqa
& & {} = \, {1 \over d}
\sum_{h=0}^{d-1}
\phi(k,k',h) \,
\phi(k,k,h) \,
\omega^{h \odot (t \oplus t')}
(\omega^{((k-1)\oplus (k'-1)) \odot h \odot h})^{\frac{1}{2}}
\delta_{((k-1) \oplus (k'-1)) \odot h,0}.
\nonumber
	\eeqa
At this point we can conclude as follows, by separating the cases 
$k \neq k'$ and $k = k'$ and then using (\ref{prop1}), (\ref{prop3}) 
and (\ref{la4}). 
	\beqa
& & \kern-40pt {1 \over d}
(1 - \delta_{k,k'})
\phi(k,k',0) \, \phi(k,k,0)
+
{1 \over d}
\delta_{k,k'}
\sum_{h=0}^{d-1}
\phi(k,k,h) \, \phi(k,k,h)
\, \omega^{h \odot (t \oplus t')} \nonumber \\
& & \kern-40pt \kern20pt {} = \, {1 \over d}
(1 - \delta_{k,k'}) + {1 \over d}
\delta_{k,k'}
\sum_{h=0}^{d-1}
\omega^{h \odot (t \oplus t')}
= \, {1 \over d} (1 - \delta_{k,k'}) + \delta_{k,k'} \delta_{t,t'}.
	\eeqa

This gives (\ref{mub}), q.e.d.


\acknowledgments

We are grateful to T. Durt for correspondence on the subject of MUB and to 
M. Lucamarini, R. Piergallini and C. Toffalori for interesting discussions
and a careful reading of the ms.


\newpage


\end{document}